# Impact of Graphene Thickness on EM Modelling of Antenna

Sasmita Dash[1,a], Christos Liaskos[2,b], Ian F. Akyildiz[3,c] and Andreas Pitsillides[1,d]

[1]Computer Science Department, University of Cyprus, Cyprus,

[2]Foundation for Research and Technology - Hellas (FORTH), Greece,

[3]School of Electrical and Computer Engineering, Georgia Institute of Technology, USA

[a]sdash001@ucy.ac.cy, [b]cliaskos@ics.forth.gr, [c]ian@ece.gatech.edu, [d]andreas.pitsillides@ucy.ac.cy

**Abstract.** This paper presents illustrative electromagnetic modelling and simulation of graphene antenna using a two-dimensional graphene sheet of zero thickness and a three-dimensional graphene slab of finite thickness. The properties of the antenna are analyzed in terms of the S11 parameter, input impedance, VSWR, radiation pattern, and frequency reconfiguration using the full-wave electromagnetic simulator. Furthermore, this work numerically studies the modelling of graphene antenna using a three-dimensional graphene thin slab and the impact of graphene slab thickness on the performance of graphene antenna.

**Introduction**

The world's first two-dimensional (2D) material graphene gained momentum for the realization of effective new devices in various fields due to its extraordinary electrical, mechanical and optical properties. Owing to unique electromagnetic (EM) properties such as tunable conductivity and slow-wave propagation, graphene attracted more attention for EM applications in a wide frequency range from microwave to X-rays. Indeed, graphene is capable of the surface plasmon polariton (SPP) propagation at terahertz (THz) frequency range [1]. Graphene plasmons enable the design of graphene plasmonic nanoantennas at THz frequency band [2-5]. However, EM simulation is an integral part of numerous research fields. A numerical simulation yields a correct and meaningful result when it implements the accurate theoretical prediction for the material properties in EM simulation. Particularly, this is the most significant concern for the modelling of graphene in computational EM simulator. Several numerical techniques can be used for graphene simulation such as finite-difference time-domain method (FDTD), finite element method (FEM), and method of moment (MoM). Based on these techniques, several commercial software packages CST, HFSS, COMSOL, and FEKO are available to model graphene antenna. Graphene can be modelled as a 2D sheet or three dimensional (3D) thin slab in commercial EM simulators.

This paper illustrates the EM modelling and simulation of the graphene 2D sheet and 3D slab antenna. Section 2 explains briefly the EM modelling and simulation of graphene using 2D sheet and 3D slab by means of computational EM simulators. Section 3 presents the EM modelling and analysis of graphene antenna using graphene as 2D graphene sheet of zero thickness and 3D graphene thin slab of non-zero thickness. Finally, the conclusion is drawn in section 4.

**Electromagnetic Modelling of Graphene**

Graphene is a single layer of $sp^2$-bonded carbon atoms packed in a hexagonal (honeycomb) lattice. The extraordinary electronic, optical and mechanical properties of graphene stimulated great research interest in numerous fields in the past few years. Graphene is a zero-overlap semimetal or zero-gap semiconductor with very high electrical conductivity. Graphene has very high electron mobility, which can theoretically reach 200,000 $cm^2$/Vs at room temperature, the current density is in the order of $10^9$ A/cm, and the velocity of fermion (electron) is the order of $10^6$ m/s, which is 300 times smaller than the velocity of light [6]. The inherent strength of graphene is another excellent property. Graphene is the strongest and lightest material ever discovered. Despite this, graphene is only 1 atom thick; it is almost transparent.

In an EM simulator, graphene is modelled as a thin conductive sheet. The graphene surface conductivity $\sigma_s$ contains contributions from intraband and interband transitions. The intraband conductivity is dominant over the interband term in the THz frequency regime. The intraband conductivity of graphene contributes a large imaginary part of conductivity, which in turn, leads to plasmonic behaviour. The intraband conductivity of graphene can be approximated by the Kubo formula [7]:

$$\sigma_{intra}(\omega,\mu_c,\tau,T) = -j \frac{e^2 \kappa_B T}{\pi \hbar^2 (\omega - j\tau^{-1})} \left[ \frac{\mu_c}{\kappa_B T} + 2\ln\left(e^{-\mu_c/\kappa_B T} + 1\right) \right] \quad (1)$$

where $e$ is the electron charge, $\omega$ is angular frequency, $K_B$ is the Boltzmann's constant, $\hbar$ is the reduced Planck's constant, $T$ is temperature, $\mu_c$ is the chemical potential, scattering rate $\Gamma$ is the electron-phonon scattering due to the carrier intraband scattering related to plasmonic loss = $1/2\tau$, and $\tau$ is the relaxation time related to impurities and defects in graphene.

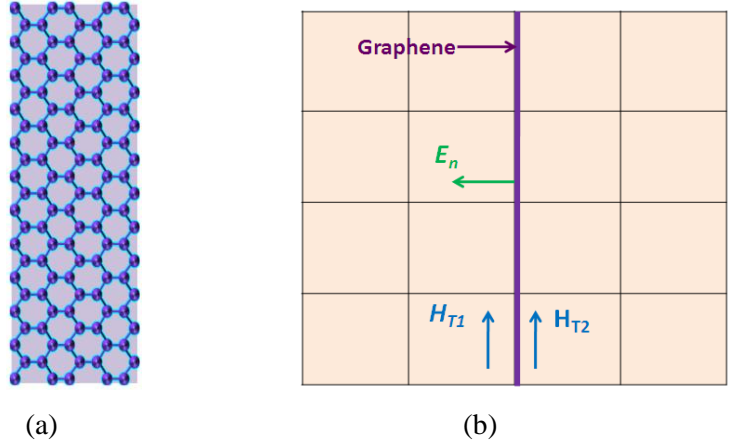

(a)        (b)

Figure 1. (a) Graphene sheet with its honeycomb lattice structure, and (b) Graphene sheet in the 2D grid.

Fig. 1 (a) shows the graphene sheet with its honeycomb lattice structure and Fig. 1 (b) shows the graphene sheet in a 2D rectangular grid, on which we have graphene as a conductive sheet with

$$\sigma_{intra} E = \hat{n} \times (H_{T1} - H_{T2}) \quad (2)$$

In most of the EM simulators, graphene is modeled as a 2D graphene sheet. The 2D graphene sheet can also be approximated by a 3D thin slab in the EM simulators. The 3D conductivity of graphene can be evaluated as

$$\sigma_{3D} = \frac{\sigma_s}{t} \quad (3)$$

where $t$ is the thickness of the graphene layer. The conductivity of graphene reduces when the graphene thickness increases [8]. Thus, the accuracy of the 3D model degrades as graphene thickness increases. To analyze the electromagnetic properties of the graphene antenna, several numerical techniques such as FDTD, FEM, and MOM can be used. Several graphene antennas have been modeled using several EM simulators based on these numerical techniques [9-15].

## Modelling and Simulation of Graphene Antenna

In this section, we illustrate aspects of the modelling of graphene antenna using a 2D graphene sheet of zero thickness and 3D graphene thin slab of non-zero thickness.

The dispersion relation allows the derivation of the resonant frequency of the graphene antenna by means of the resonance condition

$$L = m\frac{\lambda_{SPP}}{2} = m\frac{\pi}{K_{SPP}} \qquad (4)$$

where $L$ is the length of the graphene antenna, $\lambda_{SPP}$ is the SPP wavelength, and $m$ is an integer which determines the order of the resonance.

We modelled a graphene dipole antenna at 1 THz frequency. The proposed antenna structure is simulated using FEM solver Ansys HFSS 17. Fig. 2 shows the top view and cross-sectional view of the designed structure of the graphene dipole antenna. The graphene dipole antenna was modelled as a planar structure of the 2D graphene conductive sheet. The proposed antenna consists of a rectangular graphene sheet of width (W) = 10 µm and length (L) = 52 µm. We chose graphene with relaxation time $\tau$ =1 ps and temperature T = 300 K. Graphene sheet transferred on the silicon substrate. Dimension of substrate is chosen as length (Ls) = 60 µm, width (Ws) = 20 µm and height (h) = 4 µm. The source was placed in the center of the dipole antenna to excite the structure.

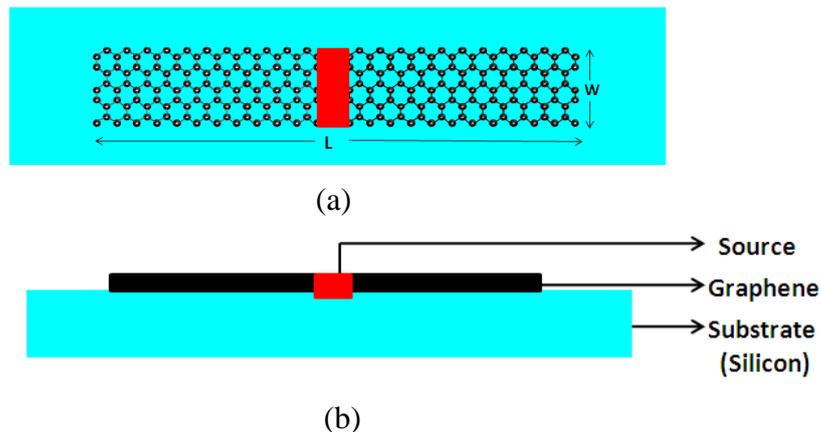

(a)

(b)

Figure 2. Proposed graphene dipole antenna (a) Top view and (b) Cross-sectional view.

## Results and Discussion

At 1 THz frequency, the SPP wavelength of the graphene dipole antenna is nearly equal to one-third of free-space wavelength ($\lambda_{SPP} \approx \lambda_0/3$). It can be seen that the graphene antenna enables high miniaturization and radiates at sub-wavelength scale. S (1, 1) parameters, input impedance, VSWR and radiation pattern of the graphene dipole antenna are shown in figure 3 (a), (b) (c) and (d) respectively. The input impedance of the antenna is nearly equal to 50 ohms. The proposed graphene antenna provides good impedance matching. VSWR value of antenna at 1 THz is equal to unity, which indicates no mismatch loss. The radiation pattern of graphene dipole antenna at 1 THz is omnidirectional in nature, similar to the metallic dipole antenna.

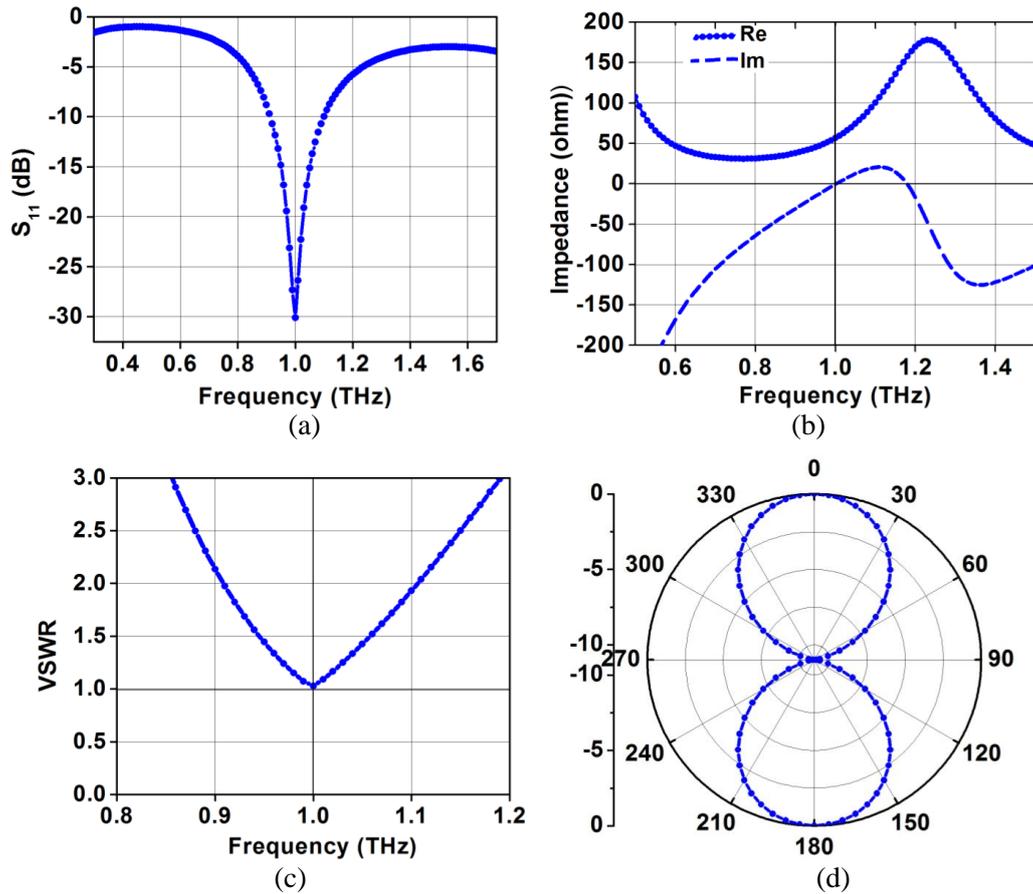

Figure 3. Performance of graphene dipole antenna (a) $S_{11}$ parameter, (b) Input impedance, (c) VSWR and (d) Radiation pattern at 1 THz.

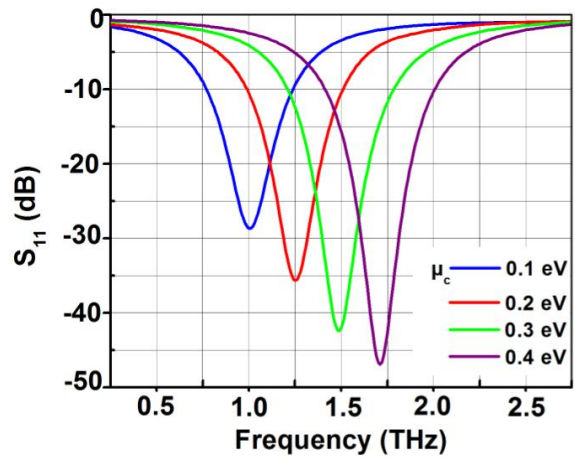

Figure 4. Frequency reconfiguration of graphene antenna.

Furthermore, frequency reconfiguration is easily achieved by varying the chemical potential parameter in the EM simulator. The conductivity of graphene increases with the increase of chemical potential, which in turn tunes the surface electromagnetic properties of graphene. Practically, the chemical potential of graphene was dynamically controlled over a wide range by applying an external gate voltage between two graphene patches of the graphene dipole antenna. In this work, Frequency reconfiguration is achieved in a frequency range 1.0 -1.75 THz with the increase of the chemical potential ($\mu_c$) of graphene from 0.1 eV to 0.4 eV (Fig. 4).

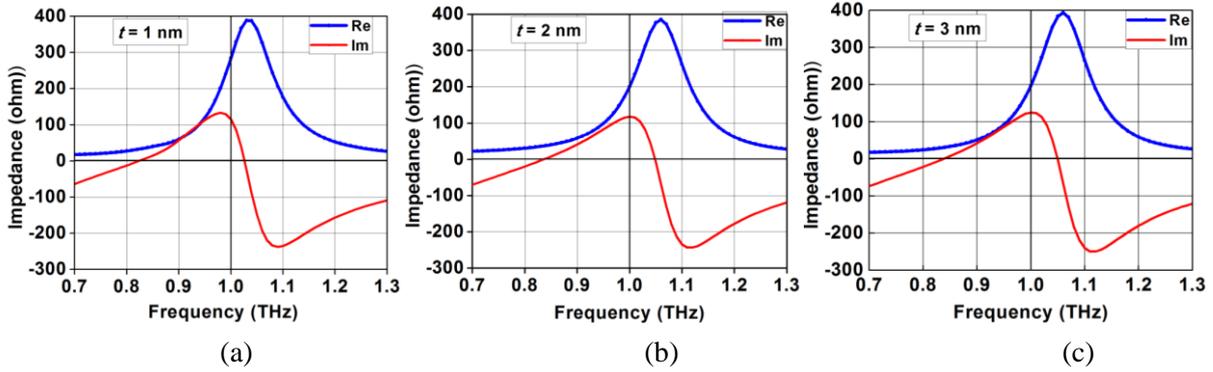

Figure 5. (a) Input impedance of graphene antenna with graphene thickness (a) 1 nm, (b) 2 nm, and (c) 3 nm.

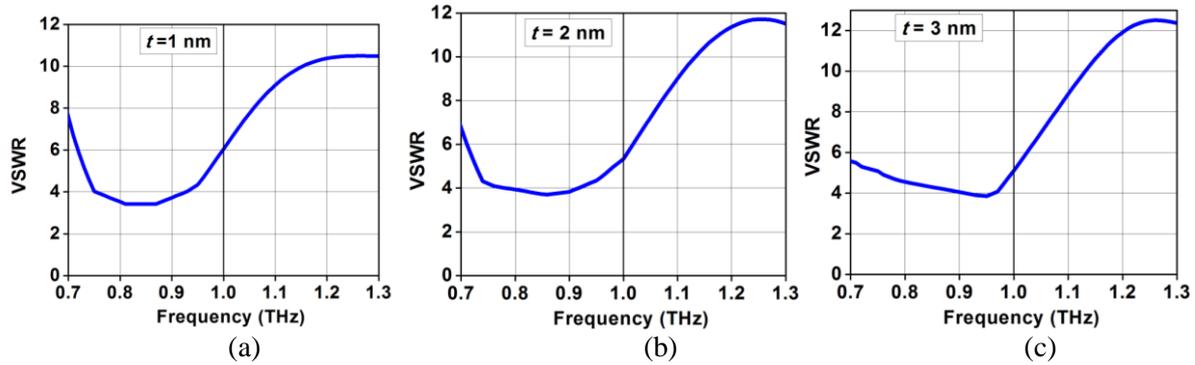

Figure 6. VSWR of graphene antenna with graphene thickness. (a) 1 nm, (b) 2 nm, and (c) 3 nm.

To illustrate the 3D aspect, the proposed antenna has been modeled using 3D graphene thin slab of variable thickness t = 1 nm, 2 nm, and 3 nm. Graphene conductivity decreases with an increase of thickness of 3D graphene slab, which in turn, reduces the current in the antenna. Thus, an antenna impedance mismatch occurs. The input impedance of proposed graphene antennas with 3D graphene slab with variable thickness is depicted in Fig.5. From these plots, it can be observed that the real part of the input impedance of the antenna at resonance 1 THz is high and for all cases, it is more than 200 ohm, whereas graphene antennas using 2D graphene sheet of zero thickness has 50-ohm input impedance. Also, the imaginary part of the input impedance of the graphene antenna using the 3D graphene slab is not zero at 1 THz. This is zero at 1.03 THz, 1.05 THz, and 1.55 THz for the thickness of 1 nm, 2nm, and 3 nm respectively. Due to impedance mismatch, VSWR of antennas for variable thickness 3D graphene slab is more than 5 at the resonant frequency 1 THz, which is shown in Fig.6. The higher value of VSWR indicates more mismatch loss.

**Conclusion**

EM modelling and simulation analysis of graphene dipole antenna using as 2D graphene sheet and 3D graphene slab has been carried out. Antenna design using a 2D conductive graphene sheet of zero thickness provided good impedance matching and no mismatch loss was found. The EM modelling and simulation of the 3D graphene slab is dependent on its thickness. The conductivity of graphene reduces when the thickness of the 3D graphene slab increases, which leads to a decrease in current density in the antenna structure and impedance mismatch. Given the illustrative simulative results, our message for the users is to take care of the selection of graphene thickness for modelling graphene antennas in the EM simulator.

**Acknowledgments**

This research is supported by the European Union via the Horizon 2020: Future Emerging Topics - Research and Innovation Action call (FETOPEN-RIA), grant EU736876, Project VISORSURF (http://www.visorsurf.eu).